# Synthesis and Stability Kinetics of Nanoporous TaC Derived from Ta Precursors

Catherine Ott[1], Vaibhav Verma[1], Adam Peters[2], Ian McCue[1]*

[1]*Department of Materials Science and Engineering, Northwestern University, Evanston, IL 60208, USA*

[2]*Department of Materials Science and Engineering, Johns Hopkins University, Baltimore, MD, 21218, USA*

*Corresponding author:* ian.mccue@northwestern.edu

**Abstract**

Ultra-high temperature ceramics (UHTCs) are promising materials for use in next-generation aerospace structures but have primarily been used as monolithic materials or coatings due to processing limitations. Here, new functionality (e.g., ablation resistance) is introduced to these materials by developing a porous form factor that can be later infiltrated with a secondary phase. This UHTC scaffold is synthesized via gas-phase carburization of nanoporous tantalum to the ultra-high-temperature ceramic, TaC. The kinetics of Ta conversion in a carburizing environment was examined over a range of temperatures to determine rate-limiting behavior and activation energy for the process. A 1-D moving interface model was constructed to predict carburization depth and compare data from the present work to that in the literature. It was found that the activation energy for carburization increases as conversion proceeds, suggesting a transition from grain boundary to bulk diffusion. Additionally, to simulate potential use cases of this nanoporous ceramic, the compositional and morphological stability was evaluated in a high temperature environment. Finally, the utility of this thermally stable porous UHTC was demonstrated through synthesis of a nanostructured composite of TaC and oxidation-resistant material, $SiO_2$.

# 1. Introduction

Ultra-high temperature ceramics (UHTCs) are typically defined as the carbides, borides and nitrides of the group IV and V transition metals which possess melting points above 3000°C [1]. Owing to strong transition-metal-to-non-metal bonding, these materials also possess high mechanical stiffness (629 GPa for TaC) [1] and thermal conductivities comparable to metals (>140 W/mK) [2]. These property combinations make UHTCs well-suited for environments with extreme temperatures, heat fluxes and mechanical loads. As a result, UHTCs have found extensive use in aerospace and re-entry vehicles. However, UHTCs face challenges throughout their lifecycle, from synthesis and sintering during processing, to oxidation and thermal shock during use, and these areas remain unresolved challenges.

A cursory examination of Ashby charts highlights the poor fracture toughness of ceramic materials. This is particularly troubling given that UHTCs are fielded in applications with frequent transient thermal conditions, which can drive failure via several distinct mechanisms [3]. One common strategy to improve their thermal resilience is the addition of reinforcement phases, such as SiC [4], but this often comes at the cost of lower service temperatures. More recently, groups have demonstrated that the fracture toughness of ceramics and other brittle materials can be substantially improved via controlled porosity [5, 6] and nanolattice form factors [5, 7]. However, these synthesis approaches are not easily scaled to commercial volumes and/or translatable to other ceramic systems.

UHTCs are commercially fielded as both coatings and monolithic materials, which impose constraints on synthesis strategies. Free-standing UHTCs are typically developed via densification of powders using techniques, such as hot isostatic pressing or spark plasma sintering, which utilize high temperatures for densification and high pressures for restricting grain growth. On the other hand, UHTC coatings are typically deposited via thermal spray, where powders are accelerated towards a substrate at high velocities and temperatures. Nevertheless, both fabrication techniques rely on powder-based feedstock, which is often synthesized by carbothermal reduction of metal oxides [1, 8]. This process involves combining the metal oxide and a reducing agent (e.g., elemental carbon), and then heating >1500°C to produce particles of the target UHTC [1]. Carbothermal reduction is economical and scalable, though it is limited by the relatively large resulting particle size (due to the high processing temperature) and high concentration of impurities (>0.1 wt. % of metals, O and C) [8].

Gas carburization is another commercial process to produce carbide materials, but it focuses on converting the surface of a metal to improve its surface hardness, wear resistance, and corrosion resistance. Carburization is an extensively studied and often-used practice in the steel industry, as it is a relatively inexpensive method for improving the properties of low-carbon steels for aggressive service conditions. The carburization process has expanded to other material systems, including refractory metals. For instance, tantalum is carburized for use in nuclear material pyrochemistry and pyrometallurgy [9], and the biomedical industry carburizes tantalum implants to improve service lifetime.

Despite its industrial uses, gas carburization is limited to surface treatments because it is a diffusion limited process – i.e., the kinetics are rate-limited by carbon transport through the carbide phase – and thus is not viable to fabricate bulk materials. One way to overcome this processing limitation, while simultaneously increasing thermal resilience, is to gas carburize porous metal powders such as those shown in Figure 1. UHTCs are comprised of refractory metals – such as Hf, Nb, Ta, Ti, V and Zr – which are all commercially available as nodular powder produced via a de-hydriding process but can also be fabricated in a more tunable form factor (i.e., with a defined morphology and characteristic length scale) using liquid metal dealloying [10-13]. Carburization of nanoscale metal powders can potentially produce nanostructured ceramics in a scalable manner, but the stability of the porous structure needs to be carefully balanced with the gas-phase conversion kinetics. This manuscript focuses on Ta because it can be easily acquired with nanoscale features and its monocarbide possesses one of the highest melting points of any known material.

Commercially, Ta is carburized at high temperatures (1600-2400°C) to drive the reaction as quickly as possible. At these temperatures, the diffusion distance of C through TaC is of order 10 μm in 5 minutes [14-19]. However, as noted in previous studies on nanoporous Ta [12], these temperatures could drive ligament coarsening before they are converted to TaC, ruining any benefits of nanoscale features. As a result, lower conversion temperatures (<1200°C) are required. Although methane dissociates readily on surfaces above 700°C [20], there was no literature evidence that Ta carburization can be carried out at temperatures less than 1200°C and what transport mechanisms might be operative for C.

In this work, we report on the low-temperature gas carburization of nanoscale Ta powders in a methane atmosphere to create nanoporous TaC. The phase transformation kinetics are evaluated as a function of time and temperature, and the activation energy of carburization is analyzed. A moving-boundary phase transformation model is constructed to better understand the evolving composition profile and compared against literature data at much higher temperatures and conversion depths. The viability of this material for aerospace applications is evaluated through exposure at high temperatures to assess phase and morphological stability. Lastly, to demonstrate proposed utility of a UHTC scaffold, the porous material is infiltrated with a secondary phase, leading to development of a UHTC-based composite. The secondary phase incorporated here is $SiO_2$, which is often used to improve the oxidation resistance of structural materials [21-23].

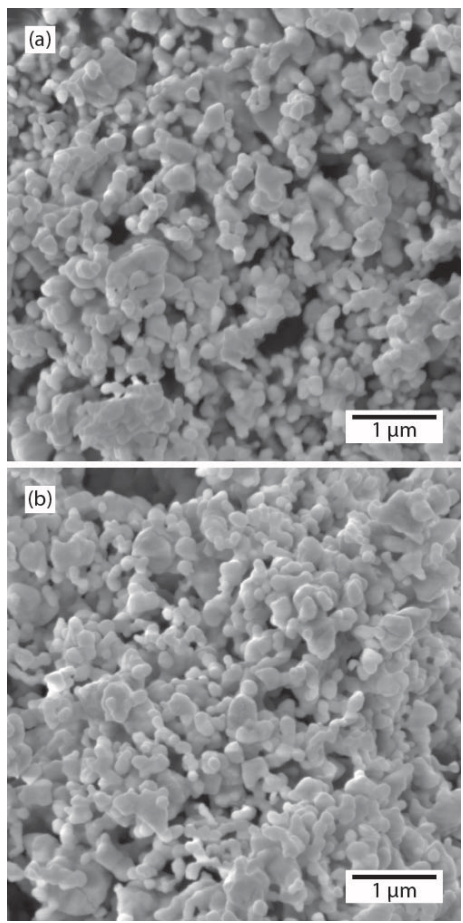

Figure 1. Nanoporous powders examined in this study. (a) Ta produced via a dehydriding process. (b) TaC produced via gas carburization at 850°C for 1 hour. The length scale of the nodular powder before and after conversion is ~190 nm.

# 2. Experimental Methods

## 2.1. Carburization and thermal annealing

### 2.1.1. Gas carburization via reaction with methane-containing gas

Gas carburization was performed on Ta powder (325 mesh, Thermo Fisher Scientific). Approximately 20 mg of Ta powder was placed in a quartz combustion boat (MTI) and spread across as a thin layer. The quartz boat was then placed in an alumina crucible (Accumet Materials). The alumina crucible was then placed in a furnace (MTI). Mass flow controllers (Alicat Scientific) were used on all gas lines to control inert and reactive gas flow.

Before heating, the furnace and all gas lines were purged at 250 sccm each of 5% $H_2$/Bal. Ar, 10% $CH_4$/Bal. Ar, and UHP Ar for 15 minutes, which was followed by another purge with 500 sccm with 5% $H_2$/Bal. Ar for 15 minutes, all at room temperature. The sample was heated to the carburization temperature at a constant heating rate of 5°C/min, with a 30-minute hold at 200°C to protect the heating elements;

heating was performed under 500 sccm 5% $H_2$/Bal. Ar. Once the carburization temperature was reached, the flowing gas was switched to 500 sccm 10% $CH_4$/Bal. Ar. The furnace was held at temperature and the reactive gas flow was maintained at 500 sccm for the specified carburization time. After the carburization time was over, the flowing gas was switched back to 500 sccm 5% $H_2$/Bal. Ar, and the furnace was cooled to 200°C at a constant cooling rate of 5°C/min. Once the furnace had cooled to 200°C, the flowing gas was changed to 100 sccm and the furnace was allowed to cool naturally until the sample was removed.

## 2.1.2. High temperature annealing of carburized powder

Once the initial gas conversion reactions were completed and characterized, the samples were subsequently annealed. When annealing for examination of the phase composition evolution, a heat treatment reaching 1400°C for 30 minutes was carried out. The samples were placed in small boats of Ta foil (Thermo Scientific), which were then placed in an alumina crucible (Accumet Materials) along with a piece of Ti sponge (Thermo scientific). Before heating, the furnace (MTI) was purged with 500 sccm of UHP Ar for 30 minutes at room temperature. The sample was heated to 1400°C at a constant heating rate of 5°C/min, with a 30-minute hold at 200°C, then held at 1400°C for 30 minutes before cooling at 5°C/min to 200°C, all while flowing 500 sccm UHP Ar.

When annealing for examination of the morphological stability, a heat treatment reaching 1700°C for 30 minutes was carried out. Before heating, the furnace (Across International) was purged with 1 scfh of UHP Ar for 30 minutes at room temperature. The sample was heated to 1200°C at a constant heating rate of 10°C/min, with a 30-minute hold at 200°C, then heated to 1700°C at a heating rate of 5°C/min and held for 30 minutes before cooling at 5°C/min to 200°C, all while flowing 1 scfh UHP Ar.

## 2.2. Phase and microstructural characterization

After carburization and the secondary heat treatments, the carbide conversion was quantified via X-ray diffraction (XRD) (Rigaku Ultima, Cu K source ($\lambda$ = 1.5406 Å)). Rietveld refinement was performed in the Match! XRD software (Crystal Impact) to quantify the phase fractions. After examining diffraction patterns collected from carburized powder samples, the phases present were Ta, TaC, $Ta_2C$, and trace amounts of $Ta_2O_5$. These phases were selected for each powder pattern to perform Rietveld refinement. The experimental data diffraction patterns were refined until the R-value and $\chi^2$ values were below 10.

Scanning electron microscopy (SEM) (FEI Quanta 650) was performed on metallic Ta powder, the carbide powder after low temperature gas carburization, and after a high temperature anneal to track the evolution of the powder particle size and morphology.

## 2.3. Kinetic analysis

### 2.3.1. Isoconversional analysis (from [21])

To gain insight into the kinetic parameters of TaC formation during gas carburization, an integral isoconversional method was employed. Isoconversional methods are rooted in the assumption that the reaction of interest behaves with single-step kinetics according to

$$\frac{d\alpha}{dt} = k(T)f(\alpha) = A \exp\left(-\frac{E}{RT}\right)f(\alpha), \tag{1}$$

where $\alpha$ is the degree of conversion, which in this case is fraction of TaC phase present; $t$ is time; $k(T)$ is the Arrhenius equation; and $f(\alpha)$ is the reaction model that describes the process. $k(T)$ is defined in terms of $A$, the pre-exponential factor; $E$, the activation energy; $R$, the universal gas constant; and $T$, the temperature. Additional details on isoconversional analysis may be found in Appendix A and in Ref. [21].

### 2.3.2. Kernel smoothing method

The collected conversion data, including the replicates at certain times and temperatures, were used to extract kinetic information with isoconversional kinetic analysis. The process of interest was conversion to TaC from Ta through gas phase carburization, so the extent of conversion $\alpha$ was defined as the phase fraction of TaC present; the carburization process will introduce a gradient in the C composition within the TaC, which could cause local lattice distortions, but this effect would be challenging to extract with our refinement. Isoconversional analysis requires comparison of the time to the same amount of conversion across different temperatures. However, achieving the exact same values of $\alpha$ across all temperatures is nearly impossible due to experimental error (surface area exposed to gas from hand packing, thermocouple variability, powder morphology, etc.). Therefore, a method of approximating the reaction time to some value of $\alpha$ for the temperatures studied was derived.

To approximate the amount of conversion at any value of $\alpha$, a kernel regression was performed on all the experimental conversion data collected at a given temperature. Kernel smoothers estimate a function by fitting a simple model, or kernel (weighting function) $K_\lambda(x_0, x_i)$, at each query point $x_0$ using the observed data points, $x_i$, around it, where data points closer to $x_0$ contribute a heavier weight. The kernel smoothing of the experimental conversion data was performed in Python with the KernelReg function in the statsmodels package; the local linear estimator, least-squares cross-validation bandwidth selection method, and a Gaussian kernel were used. The functions resulting from the kernel regression for each set of data per temperature, $T_i$, were used to extract the values of $t_{\alpha,i}$ corresponding to some $\alpha$ such that isoconversional analysis could be performed to determine the activation energy throughout the carburization process.

## 2.4. Diffusion modelling

A moving interface finite difference model was employed to predict TaC formation resulting from carbon diffusion into Ta [25]. Our model was based on a published gas-diffusion model [26], which itself was based on a model for bonding [27]. The method involves a forward Euler finite difference scheme with a fixed grid, where Lagrange interpolating polynomials allow for interface motion between discrete grid points. The formulation begins with Fick's second law, which governs the diffusion in the initial phase and the phase that forms from gas diffusion.

$$\frac{\partial c^p}{\partial t} = D^p * \nabla^2 c^p,$$ (2)

where $c^p$ is the concentration of the diffusing species (in this case, carbon) in phase $p$; $p$ can be $\alpha$, corresponding to the phase that forms from the diffusing gas species (TaC), or $\beta$, corresponding to the initial or interior phase (Ta/Ta$_2$C); $D^p$ is the diffusivity in phase $p$; and $t$ is time.

Fick's second law in 1D can then be written

$$\frac{\partial c^p}{\partial t} = D^p * \left[ \frac{\partial^2 c^p}{\partial r^2} + \frac{n}{r} * \frac{\partial c^p}{\partial r} \right],$$ (3)

where $n = 0, 1, 2$ denotes the geometry of the system (plane sheet, infinite cylinder and sphere, respectively); and it is assumed that diffusion is 1D along the direction $r$, which is normal to the surface. A mass balance condition at the interface must be met, given by

$$\frac{\partial s}{\partial t} = \frac{D^\beta * \frac{\partial c^\beta}{\partial r}\big|_{r=s} - D^\alpha * \frac{\partial c^\alpha}{\partial r}\big|_{r=s}}{c_s^\alpha - c_s^\beta},$$ (4)

where $s$ is the spatial position of the interface; the concentration of diffusing species in each phase at the interface is given by $c_s^\alpha$ and $c_s^\beta$. In the present case, $c_s^\alpha$ is the minimum solubility of C in TaC, and $c_s^\beta$ is the maximum solubility of C in Ta$_2$C, both taken from the phase diagram at the temperature of interest. The surface is exposed to a fixed concentration $c_g$ of the diffusing species.

The finite difference approximation is implemented by first discretizing the total diffusion distance, a, into a fixed amount of grid points. The interface is allowed to float between grid points, and its motion is calculated with polynomial interpolations with nearby grid points to get the spatial concentration gradient. Time steps are determined by the size of the smallest grid spacing (the distance between the interface and the nearest grid point), and the diffusivities in each phase. More information on the finite difference formulation and progression of the simulation can be found in Appendix B.

## 2.5. Development of SiO$_2$-infiltrated UHTC composites via Sol-gel processing

The process for synthesis of the Ta/TaC and SiO$_2$ composite involves three major steps: formation of the colloidal SiO$_2$ sol (hydrolysis and condensation), conversion of the sol to gel (gelation), and processing into the final dense ceramic [28]. First, tetraethyl orthosilicate (TEOS) (Thermo Fisher) was mixed with DI water; meanwhile, in a separate beaker, DI water and HNO$_3$ (Thermo Fisher) were combined. Then, the diluted HNO$_3$ was added to the TEOS solution, and mixed under magnetic stirring, leading to formation of a transparent silica-based sol. At this point, the nanoporous TaC powder was added to the sol under continuous stirring. The sol containing TaC particles was then subjected to ultrasonication using a bath sonicator (Branson) to achieve uniform dispersion of the particles. After sonication, the sol was again placed upon the stir plate and converted into gel by drop-wise addition of ammonium hydroxide (NH$_4$OH) (Thermo Fisher) under continuous stirring. The gel was allowed to set overnight, after which it was moved to a 110°C oven (Across International). The dried gel was ground manually using a mortar and pestle and sieved (325 mesh). The ground and sieved powder was calcined at 400°C in air in a box furnace (MTI).

The calcined powder was cold pressed in a cylindrical pellet die (Across International) to a stress of 300 MPa. As a binder for cold pressing, 2 wt.% isopropanol was mixed with the powder by manual grinding. Finally, the pellet was placed in an alumina crucible (Accumet Materials) and sintered under flowing Ar at 1500°C for 5 hours.

# 3. Results and Discussion

## 3.1. Phase evolution during low-temperature carburization

Although there was no literature evidence that Ta could be carburized at temperatures below 1250°C, a progressive transformation from Ta to $Ta_2C$ and TaC was found at all temperature above 700°C. The diffraction results presented in Figure 2 illustrate this conversion process at 850°C.

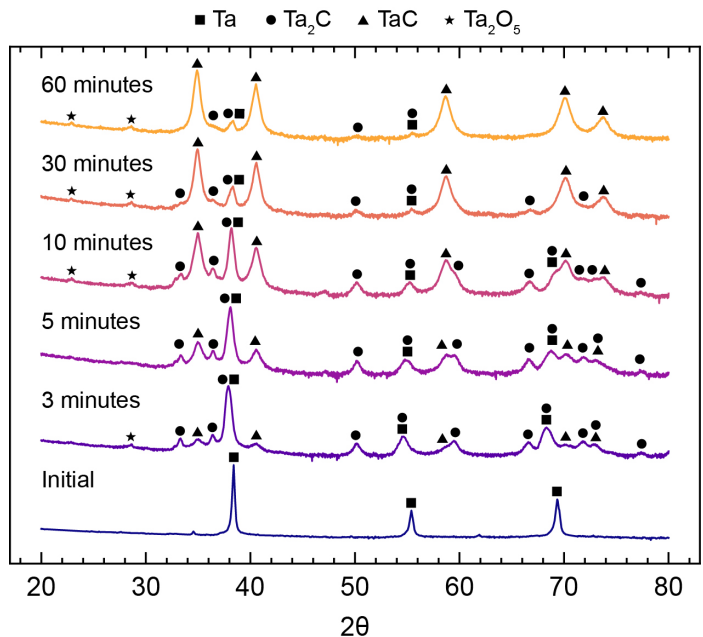

*Figure 2. Diffraction patterns for Ta carburized at 850°C for different times showing the progressive transformation from metallic Ta to primarily TaC. In line with the Ta-C phase diagram, small fractions of $Ta_2C$ appear during intermediate carburization steps, and a small fraction of $Ta_2O_5$ is measured because it is not reduced by C at these temperatures.*

Individual phase fractions have been extracted from the diffraction results and all the temperature/time data are presented in Figure 3. Ta is converted to a carbide phase for all the temperatures in this study, except for 600°C which shows little-to-no conversion even after an 8-hour isothermal hold. As expected for a thermally activated process, Ta is consumed more rapidly with increasing temperature. Lower temperatures (700-750°C) require hours, whereas the highest temperature (900°C) requires a few minutes. The TaC phase fraction is nearly inversely proportional to the Ta phase fraction, with a small fraction of $Ta_2C$ present. The conversion process appears to follow the Ta-C binary phase diagram at these temperatures, so it is expected to observe $Ta_2C$ as an intermediate between Ta and TaC. However, once the Ta is depleted, $Ta_2C$ is consumed and converted into the monocarbide phase.

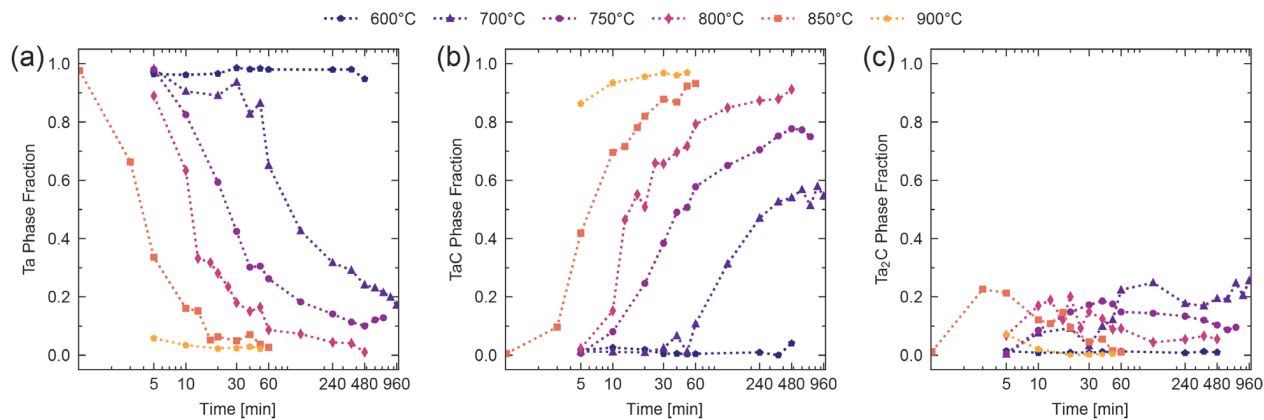

*Figure 3. Phase fraction of (a) Ta, (b) TaC, and (c) Ta₂C versus time during methane-mediated gas carburization for several reaction temperatures. Ta depletion and TaC formation closely mirror each other, following decreasing and increasing S-shaped curves, respectively. There is some Ta₂C formation in middle stages of conversion, reaching up to 30 mol % Ta₂C, but this phase depletes at high amounts of TaC.*

The results presented in Figure 3 are of a statistically significant powder quantity, but the uncertainty associated with each datapoint should be quantified before carrying out a kinetic analysis. There are several sources of error in these tests, including: thermocouple accuracy, uncertainties in phase fraction extraction from diffraction patterns, and variations in ligament diameter. To assess the effect of this variability on conversion, carburization runs were repeated for conditions where the TaC phase fraction fell between 10 and 90%. The results of duplicate runs are presented in Figure 4. The average variation for a given run was ±10 mol %, but the curves do not overlap. This lack of overlap enables the conversion kinetics to be assessed across the entire reaction.

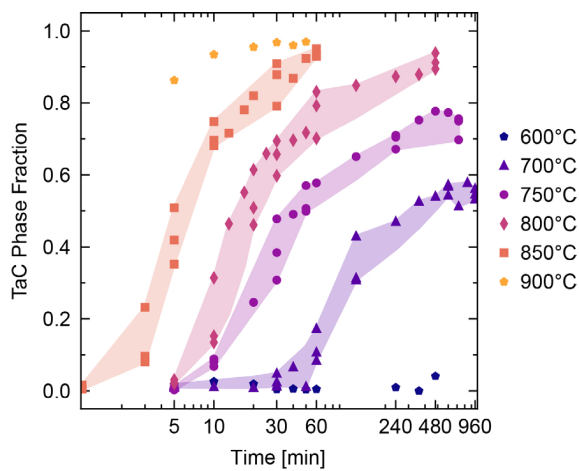

*Figure 4. TaC phase fraction versus time, including repeatability experiments for critical test conditions. Two additional tests were conducted for most temperature-time combinations when the TaC fraction fell between 10 and 90 mol%.*

## 3.2. Kinetic analysis of Ta carburization

### 3.2.1. Modeling the motion of the TaC interface with a finite difference moving boundary model

To better understand and predict carburization depth, a 1-D moving-boundary phase transformation model was constructed. Applying this model to the Ta/C system, an interface is established between the Ta$_2$C and TaC phase, and it is assumed that transport in Ta is negligible due to C having a much higher diffusivity in its lattice [14-19]. Solubility limits of C in Ta$_2$C and TaC were taken from the phase diagram at each of the temperatures studied.

In order to evaluate the efficacy of this model, two papers from the literature on Ta carburization were selected, and the reported pre-exponential factors and activation energies for diffusion of C in Ta$_2$C and TaC were used as inputs [26, 27]. The simulation was run until a representative TaC interface location was reached, and the results and the corresponding data from literature are shown in Figure 5. The moving interface model agrees reasonably well with the literature data, capturing the curvature of the data with respect to time and showing similar temperature trends. There are a few obvious outliers, such as 2300°C in both data sets and the trends for Resnick et al.'s data at 2500°C. However, this literature data was collected in the 1960s using optical measurements and likely have other experimental errors that add uncertainty to each datapoint. Nevertheless, the comparison in Figure 5 provides confidence that this model will help glean insights into the present study, where the carburization is at a smaller size scale and greater relative depth.

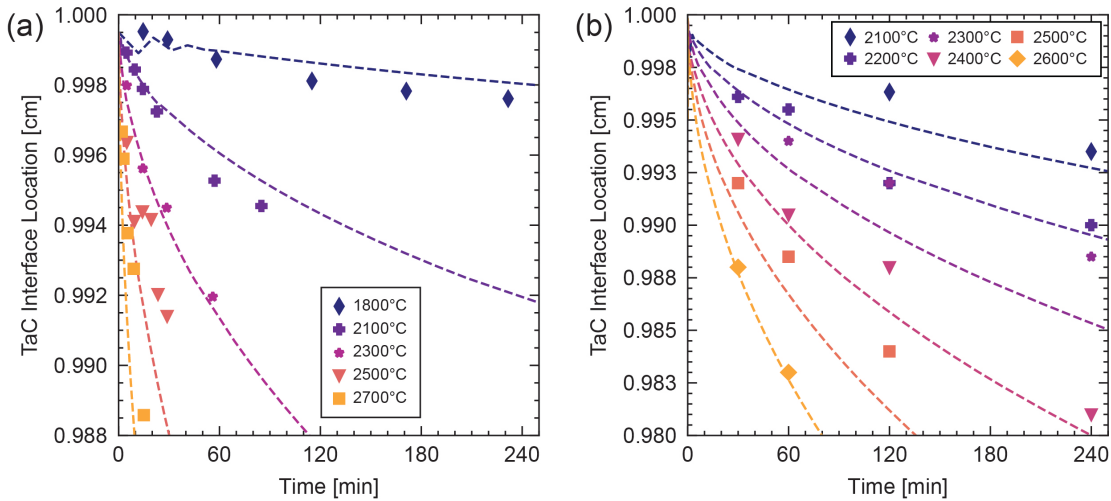

*Figure 5. Comparison of the moving interface model to experimental literature data for C diffusion into Ta. Dashed lines represent the results from the simulation, and points represent the literature data. (a) Comparison to data from Resnick et al., using the diffusion parameters reported in their study [30]. The instability in interface location for the 1800°C simulation can be attributed to initialization of the boundary away from the location where the fluxes on each side of the boundary are equal. The system must equilibrate to the flux-balance location, after which the boundary will begin to move. (b) Comparison to data from Brizes, using the diffusion parameters reported in their study [29].*

### 3.2.2. Kinetic analysis of nanoscale carburization

To compare the as-developed model with collected powder conversion data, the TaC phase fraction was transformed to a carburization distance assuming spherical particles and a shell of TaC (see Figure 4). As an initial trial, diffusion parameters ($D_0$ and E) from the literature were input in the model. However, little agreement was found with our powder carburization data across the entire range of conversion. Based on this finding, it was determined that other transport mechanisms could be at play, either due to the low reaction temperature or nanoscale features, and new diffusion parameters need to be used as input.

Diffusion parameters were calculated by performing isoconversional analysis on the kernel regression curves for low-temperature carburization of nanoporous Ta, Figure 6a. These data are plotted in Figure 6b, where each line corresponds to a different TaC conversion fraction $\alpha$. The activation energy was then extracted from the slope of each line and shown in Figure 6c for two different temperature ranges; the R-squared and standard error for the linear regressions are given in Table 1 and Table 2. The conversion data at 700°C in Figure 6a does not reach a conversion fraction $\alpha > 0.5$ for an experimentally viable timescale, and thus the orange triangles in Figure 6c do not cover the entire conversion range. However, a more complete trend versus $\alpha$ is shown in blue circles, where the activation energy is determined using 750, 800, and 850°C.

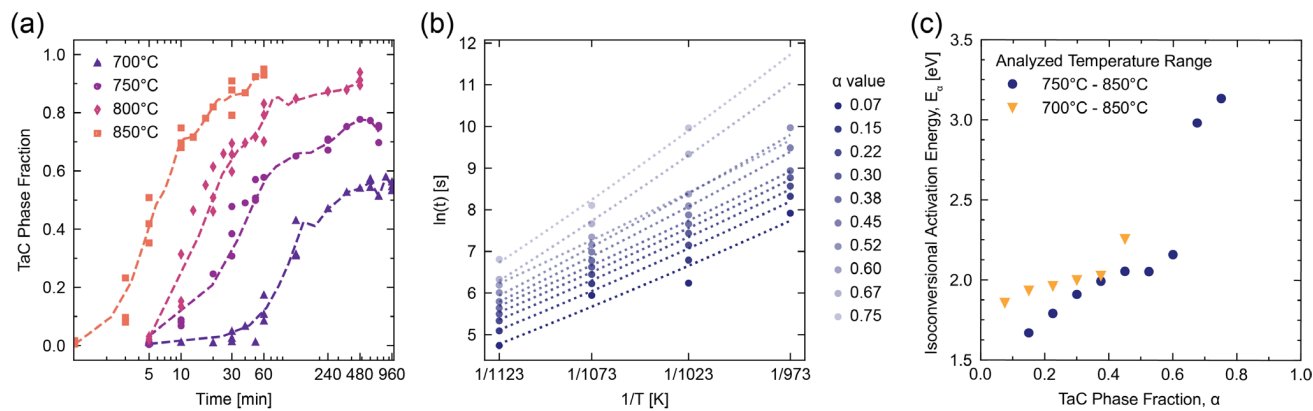

*Figure 6. Plots relevant to the kinetic analysis of the ceramic conversion process. (a) The TaC phase fraction data presented with the kernel regression of the data used to extract a single, representative conversion time for each $\alpha$. (b) Arrhenius plot of isoconversional TaC formation from low conversion percentages (dark blue) to high conversion percentages (light blue). (c) Activation energies for each amount of conversion extracted from the left plot. An $E_\alpha$ value was extracted when there were data points available, so at higher amounts of conversion, only data from conversion at 750°C-850°C could be used. The regression analysis has some error, but the mean value increases with conversion.*

The activation energies from our experiments overlap with the experimentally reported range for carbon diffusion in TaC and $Ta_2C$, which spans from 2.07-4.99 eV [19, 26-29]. Isoconversional analysis starts with the assumption that the reaction in question can be approximated by single-step kinetics. If the $E_\alpha$ value found is independent of $\alpha$, it is reasonable to conclude that the reaction can indeed be approximated by a single step. However, a variable $E_\alpha$ suggests either a multistep reaction or change in transport mechanism. While $Ta_2C$ is present as an intermediate phase, even at the initial stages of carburization (Figure 4c), it never reaches more than ~30 mol % and then decreases, giving way for TaC formation. Thus, it seems reasonable to conclude that $Ta_2C$ phase formation and depletion are not rate-limiting. It is also the opinion of the community to start evaluation of kinetic parameters with this assumption, using the results of the $E_\alpha$ versus $\alpha$ dependence to gain insights into the process [21, 30].

### 3.2.3. Applying the model to late-stage carburization

Examining the trends in Figure 6c, there is a sharp increase in activation energy at $\alpha = 0.6$ from ~2 eV to 3 eV. The activation energies for late-stage carburization in our powders are much closer to the reported literature values, so it is hypothesized that bulk diffusion of C through TaC is the dominant mechanism in this region. Furthermore, the activation energy in this region should remain mostly constant, and the moving boundary model should be applicable in this regime.

To evaluate this hypothesis, the model was initialized with a concentration profile corresponding to 60% conversion and a constant $D_0$ and E in each phase and then allowed to proceed moving the interface as before. The activation energy for C diffusion in TaC was selected to be 2.59 eV based on the $E_\alpha$ found experimentally between $\alpha = 0.6$ and 0.675. For diffusion in $Ta_2C$, the activation energy was set 0.1 eV higher than the TaC value, which is justified as $Ta_2C$ activation energy is typically higher [19, 26-29], and this condition must be met for the model curvature to agree with experiment. For $D_{C/TaC}$, $D_0$ was set at $10^{-4}$ $cm^2/s$, while for $D_{C/Ta2C}$, $D_0$ was $10^{-2}$ $cm^2/s$. The order of magnitude of these values were chosen based on $D_0$ values reported for gas carburization of bulk Ta [31]. The experimental data and the model results past 60% conversion using a "bulk diffusion" activation energy and $D_0$ are compared in Figure 7a.

Using the inputs of bulk diffusion activation energy and $D_0$, the model agrees well with experimental data for carburization beyond 60% TaC conversion. In particular, the rate of interface motion predicted by the model (i.e., the curvature of the model) aligns with the experimentally observed carburization rate at later stages. The agreement between model predictions of interface location and experimentally observed interface location is represented in Figure 7b. The proximity of these points to the identity line confirms that the current moving interface model can predict high-depth carburization behavior well in both bulk and powder form factors.

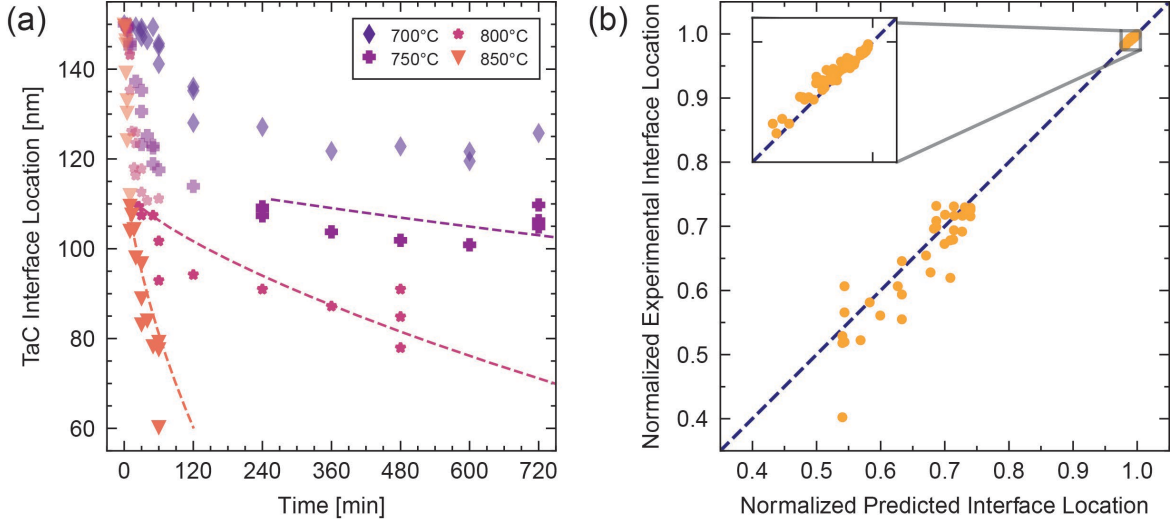

*Figure 7. (a) Comparison of the moving interface model using bulk diffusion activation energy and $D_0$ to present experimental data for TaC conversion from Ta powder for conversion amounts greater than 60 mol %. Dashed lines represent the results from the simulation, opaque points represent the experimental data corresponding to >60 mol % TaC, and transparent points are the experimental data corresponding to <60 mol % TaC. (b) Comparison of the experimental TaC interface location to the model-predicted TaC interface location for both the present powder conversion data and the bulk carburization data (called out in the inset) from [26, 27]. The identity line, y=x, represents when the experimental and predicted data agree exactly, so proximity to this line suggests better agreement between model and experiment.*

### 3.2.4. Insights into early stages of carburization at the nanoscale

It is important to note that the data in Figure 7c indicates a shift in dominant diffusion mechanism around $\alpha = 0.6$. While the orange triangles exhibit similar activation energies during the initial stage of carburization, the blue circles indicate a slight increase in the activation energy; however, as Table 1 and Table 2 show, these data points fall within each other's uncertainty. Previous studies on diffusion in tantalum have reported that diffusion in polycrystals is much faster than in single crystals, suggesting that grain boundary diffusion contributes significantly to the overall rate because bulk diffusion is energetically costly [34]. Thus, during the early stages of conversion, carbon diffusion likely occurs primarily along the grain boundaries.

Eventually, the grain boundaries become saturated with carbon and TaC has formed, so diffusion must proceed into the grain through bulk diffusion pathways. This shift from grain boundary to bulk diffusion mechanisms is reflected in the increase in activation energy as degree of conversion increases. This effect is difficult to observe in bulk conversion experiments because there is effectively an infinite source of Ta and fast-diffusing grain boundaries, but it is readily apparent in our specimens with nanoscale features and finite volumes. Furthermore, the ratio of activation energy for grain boundary and bulk diffusion reported in this work, 1.51, is comparable to literature values for carbon diffusion in TiC (1.36) and ZrC (1.54) [35].

The shift from grain boundary diffusion to bulk diffusion can be visualized by the illustration in Figure 8: as the grain boundaries saturate with C and TaC phase, diffusion and transformation further into the grains is forced, which appears as an increase in activation energy. It should be noted here that we believe that the $E_a$ extracted from the isoconversional analysis is not necessarily only the activation energy for diffusion of C in TaC. Rather, it is the energy required to move the interface itself, which may encompass multiple variables. Other changes to the structure—such as changes in site availability for diffusing atoms, grain growth at higher carburization temperatures, grain boundary structural changes with temperature, or even changing stress states from volume expansion during conversion in the nanosized powder—that occur during the carburization process could contribute to this increase in activation energy.

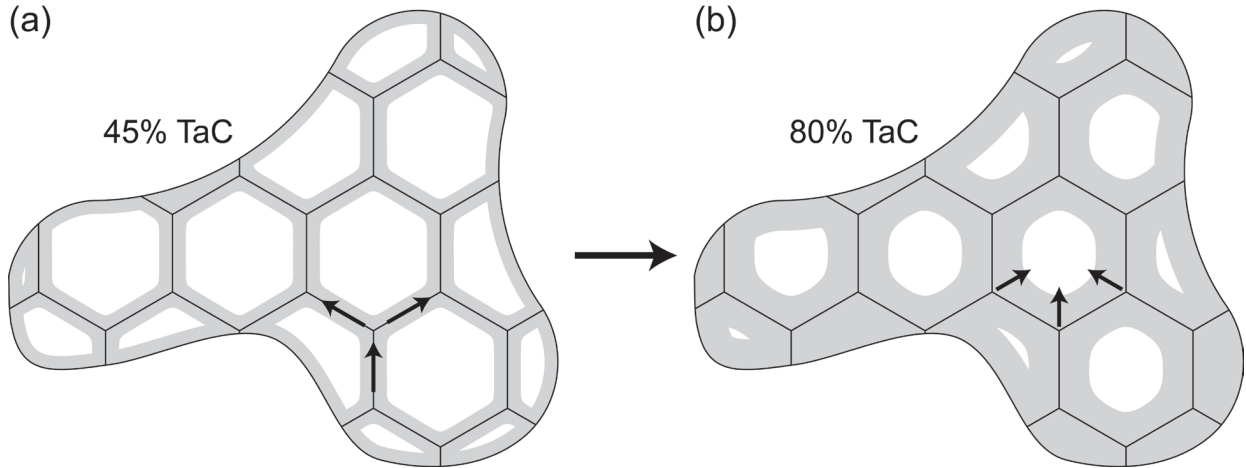

*Figure 8. Illustration of the transition from grain boundary dominated diffusion to bulk dominated diffusion that results in the increase in activation energy above α = 0.6. (a) Initially, carbon diffusion proceeds primarily through grain boundaries, resulting in TaC formation at the grain boundaries. (b) Once the grain boundaries are saturated, diffusion of carbon must proceed into the bulk, which has a higher associated activation energy.*

*Table 1. Linear regression statistics for the isoconversional activation energy ($E_\alpha$), calculated for 750°C-850°C.*

| $\alpha$ | $E_\alpha$ (eV) | $R^2$ | $S.E.$ |
|---|---|---|---|
| 0.075 | 1.47 | 0.874 | 0.559 |
| 0.15 | 1.67 | 0.953 | 0.369 |
| 0.225 | 1.79 | 0.976 | 0.280 |
| 0.3 | 1.91 | 0.985 | 0.234 |
| 0.375 | 1.99 | 0.990 | 0.199 |
| 0.45 | 2.05 | 0.988 | 0.229 |
| 0.525 | 2.05 | 0.992 | 0.183 |
| 0.6 | 2.16 | 0.997 | 0.127 |
| 0.675 | 2.98 | 0.999 | 0.112 |
| 0.75 | 3.14 | 0.994 | 0.239 |

*Table 2. Linear regression statistics for the isoconversional activation energy ($E_\alpha$), calculated for 700°C-850°C.*

| $\alpha$ | $E_\alpha$ (eV) | $R^2$ | $S.E.$ |
|---|---|---|---|

| 0.075 | 1.86 | 0.946 | 0.315 |
|---|---|---|---|
| 0.15 | 1.94 | 0.977 | 0.212 |
| 0.225 | 1.96 | 0.988 | 0.151 |
| 0.3 | 2.00 | 0.994 | 0.111 |
| 0.375 | 2.03 | 0.996 | 0.087 |
| 0.45 | 2.26 | 0.992 | 0.146 |

## 3.3. Thermal Stability of the nanoporous TaC made *via* gas carburization

### 3.3.1. Phase stability of the powders

A close examination of the results from carburization between 700-900°C, Figure 2, reveals broad peaks in the diffraction patterns, which suggests that residual strains and/or non-equilibrium phase compositions may be present. Previous studies on bulk carburization have noted equilibration of phases when carburized samples are subjected to annealing in inert gases. For instance, one study demonstrated that Ta supersaturated with carbon results in a Ta matrix decorated with $Ta_2C$ precipitates [9], and another showed that a surface TaC layer on a carburized part may decompose into $Ta_2C$ and Ta with $Ta_2C$-decorated grain boundaries [36].

During the initial conversion process, it was observed that the TaC phase fraction was nearly inversely proportional to the Ta phase fraction, with up to 30 mol % $Ta_2C$ present. In most samples containing >80 mol.% TaC after conversion, both metallic Ta and $Ta_2C$ were present, and these relative phase fractions do not agree with thermodynamic predictions given the amount of carbon incorporated. It was hypothesized that carbon diffusion at elevated temperatures may equilibrate the relative phase fraction to the ratios predicted by the phase diagram. The stability of phase composition of carburized nanoporous Ta after annealing is essential for understanding how these materials will behave in a high temperature application.

The relative phase fractions of Ta, $Ta_2C$, and TaC change substantially after annealing at 1400°C for 30 minutes. The change in phases present (and their amounts) from after carburization to after equilibration is presented in Figure 9; the relative intensity of the peaks corresponding to Ta, $Ta_2C$, and TaC shift after annealing, and the peak width decreases. The evolution of the Ta and TaC phase fractions follow a similar behavior, and small initial amounts of Ta and TaC (<50 mol.%) are lost after annealing. When only a small amount of Ta remains after carburization, annealing results in a composition of $Ta_2C$ and TaC. Similarly, when only a small amount of TaC remains, the resulting composition contains Ta and $Ta_2C$.

However, the most interesting evolution is in the $Ta_2C$ phase. Although only a maximum of 30 mol.% $Ta_2C$ forms during conversion, annealing can substantially increase this amount. For samples with sufficient carbon incorporation during the initial carburization, post-carburization annealing can equilibrate powders into almost entirely Ta$_2$C. An example of this Ta$_2$C formation is shown in Figure 9b, where after annealing 95% Ta$_2$C is present. Finally, after formation of >90 mol.% TaC during conversion, the post-annealing composition will retain the high TaC fraction.

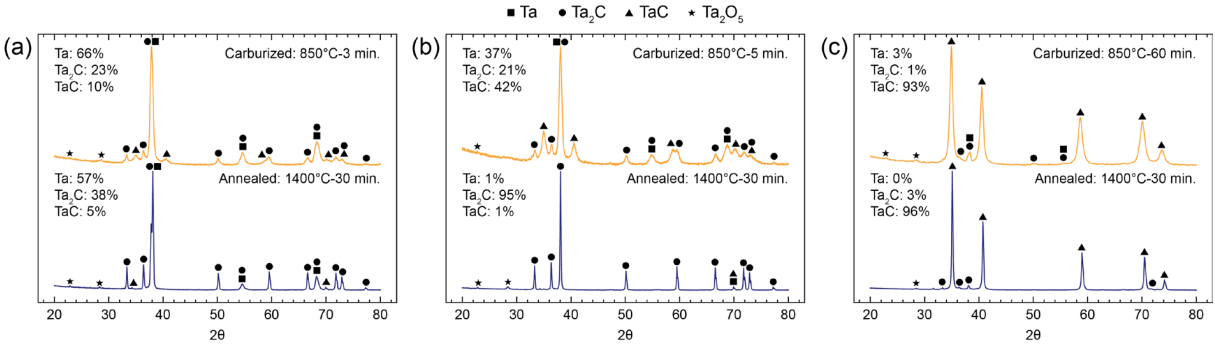

*Figure 9. X-ray diffraction patterns of converted Ta powders immediately after carburization and after annealing the same powder at 1400°C for 30 minutes. The phase fractions of Ta, Ta$_2$C, and TaC are provided to the left of each pattern. (a) Ta powder after carburization at 850°C for 3 minutes (top) and after annealing (bottom). (b) Ta powder after carburization at 850°C for 5 minutes (top) and after annealing (bottom). After only 21% Ta$_2$C was present after carburization, sufficient carbon has been incorporated to form nearly all Ta$_2$C after annealing. (c) Ta powder after carburization at 850°C for 60 minutes (top) and after annealing (bottom). No Ta phase remained after carbon was allowed to diffuse and form nearly 100 mol% TaC.*

### 3.3.2. Stability of the nanoporous morphology

High-temperature stability of the porous morphology is essential for applications where the architecture is a key design feature. The morphological coarsening of nanoporous metallic Ta powder was compared to that of Ta$_2$C (95 mol %) and TaC (90 mol %) powder made via gas carburization at 850°C for 5 minutes and 1 hour, respectively. While the converted powders are not fully Ta$_2$C/TaC, the remaining metallic Ta is likely at the center of the powder particles and the surface should have ceramic character. Thus, surface diffusion-controlled coarsening should still be governed by the phase in contact with the external environment. Both the metallic Ta and carburized TaC powders were heated to 1700°C and held for 30 minutes to allow for coarsening of their characteristic length scales. The degree of coarsening was then qualitatively evaluated via SEM imaging and is shown in Figure 10.

Comparison of the initial metallic Ta against the carbide powders (Figure 10a-c) reveals that metallic Ta does not coarsen substantially during the carburization process. The carburization process begins heating to the reaction temperature under flowing inert gas, so capillary smoothing is not competing with the conversion reaction and is the only active mechanism to consider. Surface diffusion, and thus the rate of coarsening, can be expressed with the classical form of the surface diffusion coefficient, which follows Arrhenius scaling: $D_s = D_0 \exp\left(-\frac{E_a}{k_b T}\right)$, where $D_0$ is a constant, $E_a$ is the activation energy for diffusion, $k_b$ is the Boltzmann constant, and $T$ is temperature. Tantalum has one of the highest cohesive energies among metals (4.89 eV) [37], and therefore surface diffusion is minimal at these low carburization temperatures.

After the 1700°C heat treatment the morphologies of Ta, Ta$_2$C, and TaC exhibit different feature sizes, coarsening by varying degrees. The Ta powder coarsened from ligament sizes of ~190 nm to ~3 $\mu$m, while the TaC powder maintained sub-micron ligament diameters of ~250 nm, shown in Figure 9d and Figure 9f, respectively. The coarsening of Ta$_2$C lies in between Ta and TaC, with the ligament diameter increasing to ~1 $\mu$m. These trends indicate that the energy barrier for surface diffusion in the carbide materials is larger, which is expected due to strong ionic/covalent Ta-C bonding. These strong Ta-C bonds exist in both TaC and Ta$_2$C, but the bonding character of Ta$_2$C is more metallic, and thus the activation energy is expected to be closer to that of Ta. The activation energy for diffusion is directly proportional to the cohesive energy, and TaC possesses one of the highest values (6.89 eV) [37], which supports an increased energy barrier to surface diffusion compared to metallic Ta. These results reveal that the bonding character in UHT-carbides limits mass transport along the surface and prevent substantial coarsening when used in high-temperature engineering applications.

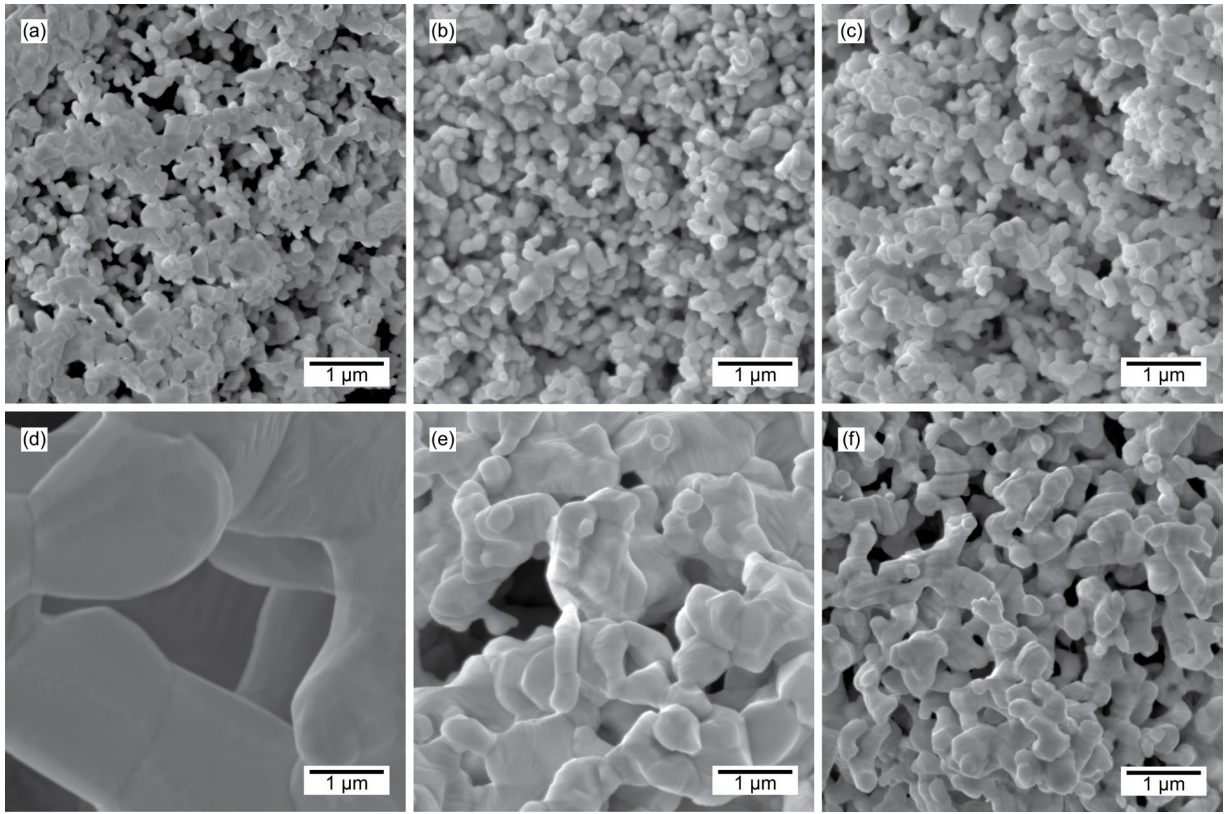

Figure 10. Scanning electron microscopy (SEM) micrographs of (a) initial nanoporous Ta, (b) powder carburized at 850°C for 5 minutes (25% Ta, 32% Ta$_2$C, 36% TaC), (c) powder carburized at 850°C for 1 hour (approximately 90 mol% TaC). (d) Ta powder from (a) after heat treatment at 1700°C for 30 minutes; (e) partially carburized powder from (b) after heat treatment at 1700°C for 30 minutes; (f) 90 mol% TaC powder from (c) after heat treatment at 1700°C for 30 minutes. There is limited coarsening during the low temperature carburization process, but exposure to 1700°C reveals substantial coarsening of the porous structure in the metallic Ta compared to the ceramic TaC.

## 3.4. Applications for nanoporous TaC

One of the motivations for fabricating a nanoporous UHTC is to use its porosity to retain molten $SiO_2$. Silica possesses one of the lowest oxygen permeabilities, and for this reason SiC is often blended into UHTCs (on the order of 20-30 vol. %) for thermal protection systems on hypersonic vehicles and re-entry systems [35-37]. However, $SiO_2$ begins to flow appreciably above 1710°C which defines its upper temperature limit [38-41]. In addition, evolution of SiO vapor at low oxygen partial pressures can cause delamination of the protective silica film at temperatures as low as 1000°C [38, 42, 43]. It is proposed in this work that a capillary network could enable higher operating temperatures for protective silica films by inhibiting aeroshearing.

The stability of a molten $SiO_2$ droplet on a surface can be estimated using expressions describing liquid motion across a surface, where droplet motion will initiate when the drag force from the shearing air overcomes the capillary force [44]. The governing equations for initiation of motion of liquid droplet can be found in Appendix C. Taking typical values for molten silica, and assuming freestream velocities above Mach 1, $SiO_2$ films should remain stable on a UHTC surface if the thickness is kept below 1 μm. For the traditional case of a UHTC blended with SiC, the $SiO_2$ scale will continuously grow as SiC oxidizes and will periodically shear off into the freestream once it reaches a critical height. However, for the case of $SiO_2$ contained within a nanoporous UHTC network, it is expected that capillary forces will prevent the glass from wicking out of the ceramic skeleton. Towards this end, a dense composite of nanoporous TaC powder (fabricated with our gas carburization method) and silica (via the sol-gel method) has been demonstrated.

The microstructure of the sintered $TaC/SiO_2$ composite is presented in Figure 11 at different magnifications. In all images, the dark phase is $SiO_2$, and the light phase is TaC. The high magnification images reveal a crack-free $SiO_2$ structure with uniform TaC distribution throughout the microstructure. In a few areas of the sample, some pores are apparent, but they are not large (< 2 μm), nor do they connect. While further optimization of the process is ongoing, the material in Figure 11 demonstrates one potential use case for a porous UHTC scaffold. However, detailed characterization of this composite's mechanical and oxidation performance is required to determine if this material can replace monolithic UHTCs.

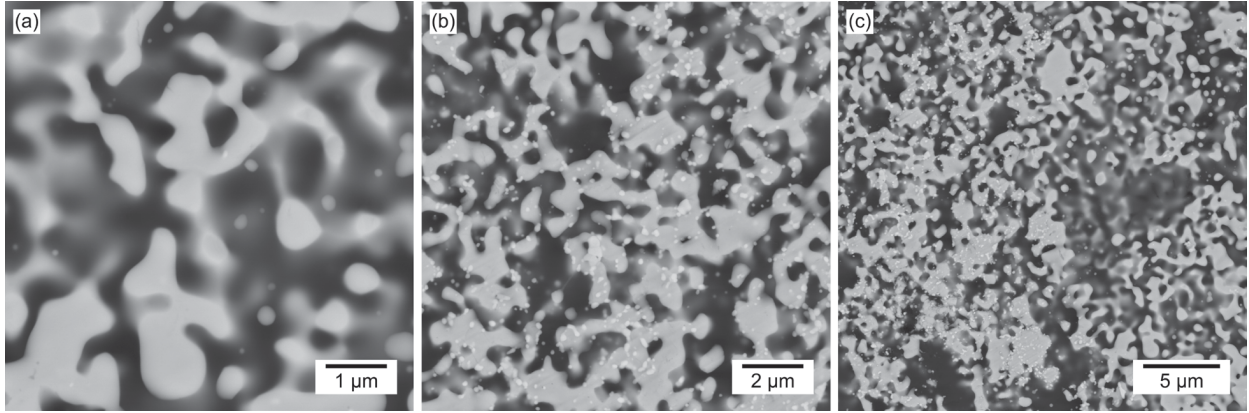

*Figure 11. SEM micrographs (backscattered electrons) of the 30 vol% TaC/70 vol% SiO₂ composite at various magnifications. The bright phase is bicontinuous TaC, where the dark phase is silica. TaC below the surface is observable where the thickness of SiO₂ is less than the electron interaction volume. (a) Micrograph collected at 20,000x magnification (b) The same specimen at 10,000x magnification. The dispersion is uniform at this length scale, and small nodules are observed on the surface. (c) The specimen at the lowest magnification, 5,000x, showing small amounts of clustering of the silica phase.*

# 4. Summary and Conclusions

This work explored the underlying mechanisms, stability, and utility of nanoporous TaC formed via a low temperature gas phase conversion process.

The kinetics of the process were examined by varying the carburization temperature (from 600°C-900°C) and time (minutes to hours), while the reactive gas, $CH_4$, was flowed over the powders. An isoconversional kinetic analysis revealed the activation energy increases with increasing TaC content, which suggested a change in diffusion mechanism at higher conversion fractions. To evaluate these mechanistic changes and predict carburization depths, a 1-D moving interface finite difference model was developed and compared to literature data for bulk Ta carburization and the current powder conversion data. The model agrees well with literature bulk carburization data and with this paper's experimental data past 60% conversion, where the sharp increase in activation energy occurs. This behavior suggests a mechanistic shift from grain-boundary/accelerated diffusion to bulk diffusion which is observed because of the geometric limitations in the nanoporous powder.

The stability of the relative phase fractions and the morphology were assessed by subjecting carburized powders to annealing heat treatments. It was found that phases will equilibrate to the ratios predicted by the phase diagram for the total incorporated carbon content. In addition, TaC shows more morphological stability than $Ta_2C$ during coarsening, which is in turn more stable than metallic Ta. This stability with increasing carbide content is attributed to a larger surface diffusion energy barrier due to the high cohesive energy in both carbides. Finally, it was demonstrated that the fabricated nanoporous TaC may be used to synthesize a nanostructured composite of $SiO_2$ and TaC, which may serve to improve the oxidation resistance of UHTCs in aggressive aerodynamic environments.

# 5. Acknowledgements

This work was supported by the Air Force Office of Scientific Research under award number FA9550-22-1-0221. This work made use of the Jerome B. Cohen X-Ray Diffraction Facility supported by the MRSEC program of the National Science Foundation (DMR-2308691) at the Materials Research Center of Northwestern University and the Soft and Hybrid Nanotechnology Experimental (SHyNE) Resource (NSF ECCS-2025633.) This work made use of the EPIC facility of Northwestern University's NUANCE Center, which has received support from the SHyNE Resource (NSF ECCS-2025633), the IIN, and Northwestern's MRSEC program (NSF DMR-2308691).

# 7. Appendices

## 7.1. Appendix A: Details on isoconversional kinetic analysis (from [21])

Isoconversional kinetic analysis methods assume that the reaction rate at some value of $\alpha$ is constant and only depends on temperature, and this can be expressed by taking the logarithm of the reaction rate and evaluating its derivative with respect to inverse temperature at a constant value of $\alpha$.

$$\left[\frac{\partial \ln\left(\frac{d\alpha}{dt}\right)}{\partial\left(\frac{1}{T}\right)}\right]_\alpha = \left[\frac{\partial \ln(A)}{\partial\left(\frac{1}{T}\right)}\right]_\alpha + \left[\frac{\partial\left(-\frac{E}{RT}\right)}{\partial\left(\frac{1}{T}\right)}\right]_\alpha + \left[\frac{\partial \ln(f(\alpha))}{\partial\left(\frac{1}{T}\right)}\right]_\alpha \tag{A.1}$$

Because $f(\alpha)$ is assumed constant at a given $\alpha$, and $A$ is a constant kinetic parameter, the equation can be reduced to the essential expression for determining $E_\alpha$, the activation energy at a specific degree of conversion.

$$\left[\frac{\partial \ln\left(\frac{d\alpha}{dt}\right)}{\partial\left(\frac{1}{T}\right)}\right]_\alpha = -\frac{E_\alpha}{R} \tag{A.2}$$

The integral form of the reaction model, $g(\alpha)$, can be found by rearranging the equation above and integrating to $\alpha$ and $t$ for an isothermal heating program, such as that used in this study.

$$g(\alpha) = \int_0^\alpha \frac{d\alpha}{dt} = A\int_0^t \exp\left(-\frac{E}{RT}\right) dt = A\exp\left(-\frac{E}{RT}\right)t \tag{A.3}$$

Combining the integral form above with the isoconversional principle, the activation energy at some $\alpha$ ($E_\alpha$) can be found from

$$\ln(t_{\alpha,i}) = \ln\left[\frac{g(\alpha)}{A_\alpha}\right] + \frac{E_\alpha}{RT_i}, \tag{A.4}$$

where $E_\alpha$ is the slope of the $\ln(t_{\alpha,i})$ versus $\frac{1}{T_i}$, where $t_{\alpha,i}$ is the time to reach conversion amount $\alpha$ at temperature $T_i$.

## 7.2. Appendix B: Details on the finite difference approximation of the moving interface

The moving interface problem defined with the equations in Section 2 are solved with the forward Euler finite difference formulation. Based on the approach from [23, 24], the concentration of the diffusing species is calculated at points on a fixed grid, while the interface is allowed to float between grid points and its motion is calculated from the mass balance condition. Because the interface floats between grid points, Lagrangian polynomial interpolation is used to approximate the spatial concentration gradient near the interface.

The diffusion direction is discretized with spacing $\delta r$, where $i$ is a given spatial node; time is discretized into variable finite time steps, where $j$ is a given time node. The concentration in phase $p$ at a discrete point in space and time is thus given by $c_{i,j}^p$.

Far from the boundary, the change in concentration at node $i$ from time step $j$ to $j + 1$ in phase $p$ is expressed as,

$$c_{i,j+1}^p = c_{i,j}^p + \delta t^p \left( \frac{\partial c^p}{\partial t^p} \right)_{i,j} \tag{B. 1}$$

The first and second space derivatives can be expressed with centered finite difference formulas:

$$\left( \frac{\partial c^p}{\partial r} \right)_{i,j} = \frac{c_{i+1}^p - c_{i-1}^p}{2\delta r} \tag{B. 2}$$

$$\left( \frac{\partial^2 c^p}{\partial r^2} \right)_{i,j} = \frac{c_{i+1}^p - 2c_i^p + c_{i-1}^p}{\delta r^2} \tag{B. 3}$$

Thus, the time derivative can be approximated as,

$$\left( \frac{\partial c^p}{\partial t^p} \right)_{i,j} = D^p \left[ \frac{c_{i+1}^p - 2c_i^p + c_{i-1}^p}{\delta r^2} + \frac{n}{r_i} \frac{c_{i+1}^p - c_{i-1}^p}{2\delta r} \right] \tag{B. 4}$$

*Near the boundary*

To approximate motion of the interface $s$, which floats between the fixed grid points, Lagrangian interpolation in the form proposed by Crank [25] and later modified by Zhou and North [27] was used. The last grid point $i$ in the $\beta$ phase (immediately preceding the interface) is named grid point $m$ – the value of $i$ assigned to $m$ will change as the interface moves into the $\beta$ phase. The value of $p$ is then defined such that the interface is located a distance $p * \delta r$ to the right of $m$. As formulated by Zhou and North [27], the motion of the interface is approximated with interpolation formulas in which the grid point closest to the interface is dropped, resulting in two cases, $p < 0.5$, where $m$ is dropped, and $p \geq 0.5$, where $m + 1$ is dropped. The second derivatives for the neighboring grid points and the first derivative for the boundary are taken from Zhou and North [27]; it should be noted that these differ than the equations reported by Schuh [26] but are the correct forms of the Lagrangian polynomials. The first derivatives for neighboring grid points (required for the cylinder and sphere geometries) are not given by Zhou and North [27], so the equations given by Schuh [26] are corrected here.

The spatial derivative approximations are:

If $p < 0.5$:

$$\left(\frac{\partial^2 c^\beta}{\partial r^2}\right)_{m-1} = \frac{2}{\delta r^2}\left[\frac{c_{m-2}^\beta}{2+p} - \frac{c_{m-1}^\beta}{1+p} + \frac{c_s^\beta}{(1+p)(2+p)}\right] \tag{B. 5}$$

$$\left(\frac{\partial c^\beta}{\partial r}\right)_{m-1} = \frac{1}{\delta r}\left[-\frac{(1+p)c_{m-2}^\beta}{2+p} + \frac{pc_{m-1}^\beta}{1+p} + \frac{c_s^\beta}{(1+p)(2+p)}\right] \tag{B. 6}$$

$$\left(\frac{\partial c^\beta}{\partial r}\right)_{boundary} = \frac{1}{\delta r}\left[\frac{(p+1)c_{m-2}^\beta}{2+p} + \frac{(2+p)c_{m-1}^\beta}{1+p} + \frac{(2p+3)c_s^\beta}{(1+p)(2+p)}\right] \tag{B. 7}$$

$$\left(\frac{\partial^2 c^\alpha}{\partial r^2}\right)_{m+1} = \frac{2}{\delta r^2}\left[\frac{c_{m+2}^\alpha}{2-p} - \frac{c_{m+1}^\alpha}{1-p} + \frac{c_s^\alpha}{(1-p)(2-p)}\right] \tag{B. 8}$$

$$\left(\frac{\partial c^\alpha}{\partial r}\right)_{m+1} = \frac{1}{\delta r}\left[\frac{(p-1)c_{m+2}^\alpha}{p-2} + \frac{pc_{m+1}^\alpha}{1-p} + \frac{c_s^\alpha}{(p-2)(p-1)}\right] \tag{B. 9}$$

$$\left(\frac{\partial c^\alpha}{\partial r}\right)_{boundary} = \frac{1}{\delta r}\left[\frac{(p-1)c_{m+2}^\alpha}{2-p} + \frac{(2-p)c_{m+1}^\alpha}{1-p} + \frac{(2p-3)c_s^\alpha}{(1-p)(2-p)}\right] \tag{B. 10}$$

If $p \geq 0.5$:

$$\left(\frac{\partial^2 c^\beta}{\partial r^2}\right)_m = \frac{2}{\delta r^2}\left[\frac{c_{m-1}^\beta}{1+p} - \frac{c_m^\beta}{p} + \frac{c_s^\beta}{p(1+p)}\right] \tag{B. 11}$$

$$\left(\frac{\partial c^\beta}{\partial r}\right)_m = \frac{1}{\delta r}\left[-\frac{pc_{m-1}^\beta}{1+p} + \frac{(p-1)c_m^\beta}{p} + \frac{c_s^\beta}{p(1+p)}\right] \tag{B. 12}$$

$$\left(\frac{\partial c^\beta}{\partial r}\right)_{boundary} = \frac{1}{\delta r}\left[\frac{pc_{m-1}^\beta}{1+p} - \frac{(1+p)c_m^\beta}{p} + \frac{(2p+1)c_s^\beta}{p(1+p)}\right] \tag{B. 13}$$

$$\left(\frac{\partial^2 c^\alpha}{\partial r^2}\right)_{m+2} = \frac{2}{\delta r^2}\left[\frac{c_{m+3}^\alpha}{3-p} - \frac{c_{m+2}^\alpha}{2-p} + \frac{c_s^\alpha}{(2-p)(3-p)}\right] \tag{B. 14}$$

$$\left(\frac{\partial c^\alpha}{\partial r}\right)_{m+2} = \frac{1}{\delta r}\left[\frac{(p-2)c_{m+3}^\alpha}{p-3} + \frac{(1-p)c_{m+2}^\alpha}{p-2} - \frac{c_s^\alpha}{(p-3)(p-2)}\right] \tag{B. 15}$$

$$\left(\frac{\partial c^\alpha}{\partial r}\right)_{boundary} = \frac{1}{\delta r}\left[\frac{(p-2)c_{m+3}^\alpha}{3-p} + \frac{(3-p)c_{m+2}^\alpha}{2-p} + \frac{(2p-5)c_s^\alpha}{(2-p)(3-p)}\right] \tag{B. 16}$$

The motion of the interface is determined from the mass balance condition and can be expressed with the boundary concentration gradients above as follows,

$$p_{i,j+1} = p_{i,j} + \frac{\delta t}{\delta r}\left[\frac{D^\beta\left(\frac{\partial c^\beta}{\partial r}\right)_{boundary} - D^\alpha\left(\frac{\partial c^\alpha}{\partial r}\right)_{boundary}}{c_s^\alpha - c_s^\beta}\right] \tag{B. 17}$$

The time step $\delta t$ is defined as the smaller of either $\delta t^\alpha$ or $\delta t^\beta$ which are given by,

$$\delta t^p \leq \frac{1}{2D^p} \Delta R^2 \tag{B. 18}$$

$\Delta R$ is the smallest grid spacing in the time step, which is around the interface. Thus, when $p < 0.5$, $\Delta R = \delta r * (1 - p)$, and when $p \geq 0.5$, $\Delta R = p * \delta r$.

*Special cases*

Special consideration must be taken when the interface is located close to the surface and close to the center. When the interface is close to the surface, the concentration profile in the $\alpha$ phase is calculated based on a linear interpolation from the surface to the interface. Similarly, when the interface is close to the center, the concentration in the $\beta$ phase is a linear interpolation from the center to the interface. Additionally, the initialization of the interface must be addressed; in the present model, the interface is initialized within the last grid spacing.

## 7.3. Appendix C: Expressions regarding a liquid droplet on a surface subjected to aeroshear.

A liquid droplet will begin to move, or "skid", across a surface when drag force on the droplet exceeds the capillary force. This condition is described with the following equations [44]:

$$F_{drag} > -F_{capillary} \tag{C. 1}$$

$$F_{capillary} = -2\sigma R_D (\cos(\theta_{rear}) - \cos(\theta_{front})) \tag{C. 2}$$

$$F_{drag} = k_1 R_D H_0 Re^{-0.25} \rho V_{air}^2, \quad k_1 = 0.35 \tag{C. 3}$$

where $\sigma$ is the surface tension of the liquid, $R_D$ is the droplet radius, $\theta$ is the contact angle of the front/rear of the droplet; $H_0$ is the height of the droplet; $Re$ is the Reynolds number; $\rho$ is the liquid density; and $V_{air}$ is the velocity of the freestream.